\documentstyle[preprint,aps,epsfig]{revtex}
\tightenlines
\begin{document}
\draft
\title{
First and second order transition in frustrated XY systems
       }
\author{D. Loison and K.D. Schotte }
\address{
Institut f\"ur Theoretische Physik, Freie Universit\"at Berlin,
Arnimallee 14, 14195 Berlin, Germany\\
Loison@physik.fu-berlin.de, Schotte@physik.fu-berlin.de
}
\maketitle
\begin{abstract}
The nature of the phase transition for the $XY$ stacked triangular
antiferromagnet (STA) is a controversial subject at present. The
field theoretical renormalization group (RG) in three dimensions
predicts a first order transition. This prediction disagrees with
Monte Carlo (MC) simulations which favor a new universality class
or a tricritical transition.
We simulate by the Monte Carlo method two models derived from the
STA by imposing the constraint of local rigidity which should have
the same critical behavior as the original model.
A strong first order transition is found.
Following Zumbach we analyze the second order transition observed
in MC studies as due to a fixed point in the complex plane.
We review the experimental results in order to clarify the different
critical behavior observed.

\end{abstract}
\vspace{1.cm}
P.A.C.S. numbers:05.70.Fh, 64.60.Cn, 75.10.-b 
\section{INTRODUCTION}

Phase transitions of frustrated spin systems have been 
extensively studied during
the last decade (for reviews see \cite{Diep2}). In particular the nature
of the phase transition of the stacked triangular antiferromagnet (STA) with
$XY$ spins interacting via nearest--neighbor bonds has been
extensively investigated \cite{Kawa88,Kawa92,PlumerXY,Loison 96}. 
At high temperatures the symmetry group of
this system is $O(2)\otimes Z_2$ whereas at low temperatures this symmetry
is completely broken. The Ising symmetry $Z_2$ has its origin in the
non collinearity of the spins in the ground state which can be classified
as chirality plus or minus. For non frustrated systems the symmetry group
in the high temperature region is simply $O(2)$ and this difference in
symmetry between frustrated and non frustrated spin systems should lead to a
different critical behavior. Bailin \cite{Bailin} and Garel \cite{Garel},
using the
renormalization group (RG), proved that there is no stable fixed point
close to space dimension $d = 4$ and they concluded that the transition
is of first order. Extending the RG technique to a $N$ component spin system,
that is using $4 - \epsilon$ expansion to first order in $\epsilon$ (two-loops)
and expanding also in $1/N$, Kawamura \cite{Kawa88}
suggested a new universality class
linked to the chirality for the transition with the exponents given by Monte
Carlo simulations \cite{Kawa92}.  
With the same technique in $d=4-\epsilon$ and in three dimensions 
to three-loops, 
it was shown later that the transition for $N = 2$
must be of first order \cite{Antonenko 94,Antonenko2}. 
Further Monte Carlo studies \cite{PlumerXY,Loison 96} have confirmed
that the exponents for STA--system (see table \ref{table5}) 
are different from the ones
of the standard $O(N)$ universality class (given in table \ref{table6}). 
Simulations seem
to favor the concept of a new chiral universality class or tricritical
behavior.
However, Plumer and Mailhot \cite{Plumer97}
used different exchange constants for
spins along the c--axis and spins in the triangular planes, so that
the hexagonal STA--system is quasi one-dimensional. They concluded that the
transition is weakly first order.
\par
There are two principal groups of magnetic materials which can be
modeled by our system.  
The first group are Hexagonal perovskites ABX$_3$,
which are quasi one dimensional systems which however order at low temperatures
and have a planar anisotropy so that the
spins are in the $XY$--plane. 
The most studied examples are CsMnBr$_3$
\cite{Mason2,Mason3,Ajiro,Gaulin,Mason1,Kadowaki1,Wang1,Weber,Goto,Deutschmann,Collins1}, 
RbMnBr$_3$ \cite{Kato1,Kato2}, CsVBr$_3$ \cite{Tanaka}, 
CsVCl$_3$ \cite{Hikokawa}, and CsCuCl$_3$ \cite{Adachi,Weber2}.
For a review see ref. \cite{Collins1}.
The results of the first four compounds are compatible with second order
transitions with exponents more or less in agreement with the MC simulations:
for example $\nu = 0.50(1)$ in MC and between 0.54(3) and 0.57(3)
experimentally (for details see Tables \ref{table1}). 
However, the specific heat
measurement of CsCuCl$_3$ indicates a cross over to first order in zero
magnetic field \cite{Weber2}.
\par
The second group are helimagnetic systems.
Since the angle between the spins can be different from 120$^{\circ}$ for
the STA--structure without changing the critical behavior\cite{Kawa90}, 
helimagnetic rare
earths (see ref. \cite{Koehler}, 
for Ho \cite{Tindall2,Tindall1,Jayasuriya1,Thurston1,Gaulin1,DuPlessis1,Eckert1,Helgesen}, 
for Dy \cite{Zochowski,Astrom,Lederman,Jayasuriya,DuPlessis2,Brits,Loh} 
and for Tb \cite{Jayasuriya3,Dietrich,Tang1,Hirota,Tang2})
could also be analyzed by the STA--model. For helimagnets the
critical behavior is quite varied (see the review for Ho and Dy 
ref. \cite{DuPlessis1}).
Essentially three types of results exist: one in favor of a $O(4)$ class
\cite{DuPlessis1,Eckert1,DuPlessis2,Brits}, 
another in favor of a new universality class
\cite{Jayasuriya1,Gaulin1,Lederman,Jayasuriya,Jayasuriya3,Dietrich,Tang1,Hirota,Tang2},
and a third class which favor first order transition 
\cite{Tindall1,Zochowski,Astrom}. See
Tables \ref{table2}-\ref{table4}
for details. With these results a definite answer cannot be
given about the order of transition. In section V we will come back to
this point.

In order to check the results of the renormalization group studies
\cite{Antonenko 94} 
we have studied the STAR and the Stiefel model \cite{Kunz}. 
The first is derived from
the STA model by imposing the constraint that in each triangle the sum of
the spins is zero at all temperatures. The modes removed are irrelevant
for the RG and the two models STA and STAR should be in the same
universality class, provided such a class exists. As is explained in the
next section the Stiefel model we use for the simulation is connected to
the STAR model. Each cell of three spins plus constraint is equivalent to a
system of dihedral, i.e.\ an ensemble of two perpendicular vectors. Neighboring
pairs of vectors interact ferromagnetically, but only vectors of the same kind.
Since these two systems have the same number of degrees of freedom, they should
belong to the same universality class. We think that the two
models are closer to the RG studies than the original stacked triangle
antiferromagnet. We can therefore check the predictions of the
RG and gain an understanding of the difficulties
one has with the results of Monte Carlo simulations and measurements
in the critical region.

In section II we present the two models. The simulations and the details
of the finite size scaling analysis for a first order transition are
explained in section III. The results are shown in section IV. Discussion
and conclusions are in section V.

\section {Model and simulation}
\subsection {The STAR model}
Starting from the stacked triangular antiferromagnet (STA)
we take the simplest Hamiltonian with one antiferromagnetic
interaction constant $J > 0$
\begin{equation}
\label{tata1}
	H = J \sum_{(ij)} {\bf S}_{i}.{\bf S}_{j} \ ,
\end{equation}
where ${\bf S}_{i}$ are  two component classical vectors of unit length.
The sum runs over all nearest neighbor pairs, that is the spin ${\bf S}_{i}$
has six nearest neighbor spins in the same $XY$--plane and two in
$Z$--direction in adjacent planes.
The ground state is characterized by a planar spin configuration with
three spins on each triangle forming a $120^{\circ}$ structure with either
positive or negative chirality (see Fig. 1). The ground state degeneracy 
is thus twofold like the Ising symmetry, in addition to the continuous
degeneracy due to global rotations.

In the RG theory one uses the concept of local rigidity which means that
the sum of three spins ${\bf S}_{1}$, ${\bf S}_{2}$, ${\bf S}_{3}$ on the
corners of a triangle is set to zero
\begin{equation}
	{\bf S}_{1}+{\bf S}_{2}+{\bf S}_{3} = {\bf 0} \ .
\label{star2}
\end{equation}
In this theory the local fluctuations violating this constraint
become modes with a gap. Thus they do not contribute to the critical
behavior and can be neglected \cite{Delamotte}.
 The constraint in (\ref{star2}) used for
an ordinary collinear antiferromagnet with two spins instead of three
eliminates also one degree of freedom so that only one is left
which means that the critical behavior of antiferromagnet is the same as
that of a ferromagnet. In our case we are left with two degrees of freedom.
We choose one spin direction and then have one more choice
for the chirality, that is the direction of the second spin could be chosen
clockwise or counter--clockwise with respect to the first spin.

In order to impose local rigidity for the MC simulation, we first
partition the lattice into interacting triangles which do not have
common corners. This can be done as follows. In each $XY$--plane
one selects in a row one "supertriangle". Then one finds two nearest 
supertriangles which do not share a common corner in the row (they are
separated by two head-up and three head-down triangles). This process is
repeated. Then one takes the next row until all rows in the $XY$--planes are
filled with triangles which do not share corners, see Fig.\thinspace 1.
All spins are then located on the supertriangles and each spin belongs to only
one supertriangle. Local rigidity  means that the three spins in each
supertriangle form a $120^{\circ}$ structure. Only in the ground state all
the supertriangle have the same orientation. At finite temperature
local rigidity means that there are no local fluctuations within a
supertriangle, but fluctuations between supertriangles are allowed.

The MC updating procedure for the state of the supertriangles is made as 
follows. At a supertriangle, we take a new random orientation for one
of its three spins; next we choose a second spin
so as to form a $\pm 120^{\circ}$ angle with the first spin. For the
orientation of the third no freedom is left.
The interaction energy between the spins of this supertriangle with the 
spin of the neighboring supertriangles is calculated in the usual way and
we follow the standard Metropolis algorithm to update one supertriangle
after the other.

We consider $L*L*L_z$ 
systems , where $L*L$ is the size of the planes, and $L_z=2 L/3$ the
number of planes. $L$ must be a multiple of three so 
that no frustration occurs because of periodic boundary conditions in
the $XY$--planes. Simulations have been done for systems sizes with
$L =12,\,18,\,24,\,30,\,36$.

The order parameter $M$ used in the calculation is
\begin{equation}
   M = {1 \over N} \,\sum_{s = 1}^3 |\,M_{s}| \ ,
\end{equation}
where $M_{s}$ $(s=1,2,3)$ is the s-th sublattice
magnetization and $N=L^2 L_z$ is the total number of the lattice sites. 
This definition corresponds to the one for the ordinary antiferromagnet
with only two sublattices. Instead of  alternating signs for the collinear case
the non collinear staggered magnetization is obtained
by making a rotation of $+120^0$ ($-120^0$) for the second (third)
magnetization before summing over the three sublattice magnetizations.

The chirality $\kappa$ is defined in the usual way
\begin{eqnarray}
\label{tata6}
{\bf \kappa}_i &  = & \frac {2}{3 \sqrt{3}} \, \Bigl{[} {\bf S}^1_i
\times {\bf S}^2_i \, + \, {\bf S}^2_i \times {\bf S}^3_i \, + \, {\bf S}^3_i
\times {\bf S}^1_i  \Bigr{]} \ ,
\\
\label{tata8}
\kappa &  = &  {1 \over N'} \,\big| \, \sum_{i} {\bf \kappa}_i \, \big| \ ,
\end{eqnarray}
where the summation is over all supertriangles and $N'=N/3$ is their number.
The chirality $\kappa_i$ of one triangle is parallel to
the $Z$-axis and equal to $\pm1$.

\subsection {The Stiefel model}
The Stiefel model can be derived from the STAR model \cite{Delamotte}.
We give the main points. In each elementary cell an orthonormal basis 
\begin{equation}
{{\bf e}_a(i)};\  a=1,2
\end{equation}
is defined, where $i$ is the superlattice index. Each spin located in the
cell can be represented in this basis
\begin{equation}
{\bf S}_i = \sum_{a} s_a(i)\,{\bf e}_a(i) \ .
\end{equation}
If we put this expression into the Hamiltonian (\ref{tata1}) we obtain
a new Hamiltonian with interactions between the orthogonal vectors
${\bf e}_a(x)$:
\begin{equation}
\label{tata2}
H = J \sum_{ij}  \Big{[} \ {\bf e}_{1}(i)\cdot{\bf e}_{1}(j) \, + \,
{\bf e}_{2}(i)\cdot {\bf e}_{2}(j) \ \Big{]} \ .
\end{equation}
The interaction can be chosen negative (or ferromagnetic) and the sum
$\sum_{ij}$ is for simplicity over nearest neighbor pairs of a simple cubic
lattice instead of a hexagonal lattice since the new spins ${\bf e}_{a}$
(see Fig. 2 and 3) are no longer frustrated. The chirality $\kappa$ for
the Stiefel model is defined as
\begin{equation}
\label{tata7}
\kappa\, = \,{1\over N}\,
\bigl|\,\mbox{$\sum_{i}$} {\bf e}_{1}(i) \times {\bf e}_{2}(i)\, \bigr| \ .
\end{equation}

The Hamiltonian (\ref{tata2}) is similar to the one of the Ashkin-Teller model
\cite{Askin}. Indeed we can give this Hamiltonian a form
which is close to it. The interaction energy of two nearest
neighbors (ij) with opposite chirality (\ref{tata7}) is zero and with 
the same chirality it is $\,2\,{\bf e}_{1}(i) {\bf e}_{1}(j)$ (see Fig. 3).
Therefore the Hamiltonian can be written as 
\begin{equation}
H=J \sum_{i,j} \, (1 + \sigma_i \sigma_j) \, {\bf S}_i {\bf S}_j 
\end{equation}
where $\sigma=\pm 1$ is an Ising spin representing the chirality 
and ${\bf S}_i$ is an $XY$--spin. In Hamiltonian of Ashkin and Teller
only Ising spins appear.

Despite the fact that the Stiefel model is extensively studied,
especially in two dimensions, no clear picture emerged.
The problem is to know whether there are two transitions, an Ising and a
Kosterlitz--Thouless transition, or only one transition of a new type
\cite{Lee1,Loison 98}. In three dimensions it has been shown 
that there is only one transition \cite{Kawa88,Kawa92,Loison 96,PlumerXY}.
Here the problem is to determine the order of the transition.

The procedure of MC procedure is as follows.
At each site one takes a new random orientation for the first
vector and chooses for the second vector a perpendicular direction
(we have two choices: $\pm90^0$, the Ising degrees of freedom).
We have two degrees of freedom the same number as for the STAR model.
The interaction energy between this dihedral and its neighbors 
is calculated. If it is lower than the
energy of the old state, the new state is
accepted. Otherwise, it is accepted only with a probability according to
the standard Metropolis algorithm.
It is possible to use a cluster MC algorithm \cite{Kunz}, but in the case
of a strong first order transition there is no reduction of the critical
slowing down \cite{Janke 93}. Periodic boundary conditions are used.

Systems with $L=12,\,15,\,18,\,21,\,24$ have been simulated.
To compare with the size $L$ 
of the STA or the STAR model, we must multiply $L$ by $\sqrt{3}$. One 
supertriangle or triangle
contains three spins and is represented
by one site in the Stiefel model. So we obtain equivalent sizes 
from 20 to 40.

The order parameter $M$ for this model is
\begin{equation}
M = {1 \over 2 N}\, \sum_{s=1}^2 \,\big|\,M_{s}\big| \ ,
\end{equation}
where $M_{s}$ $(s=1,2)$ is the
magnetization for the vectors ${\bf e}_\alpha$ over all sites
and $N=L^3$ is the total number of sites.

\subsection{ Definitions, histogram methods and finite-size scaling }     

We use in this work the histogram MC technique
developed by Ferrenberg and Swendsen \cite{Ferren88,Ferren89}.
The histogram for the energy $P_T(E)$ 
is very useful for identifing a first order transition.
Also the data obtained by simulation at $T_{0}$ 
can be used to obtain thermodynamic quantities at temperature $T$ close
to $T_{0}$.
Since the energy spectrum of a Heisenberg spin system is continuous, the
data list obtained from a simulation is basically a histogram with one entry
per energy value. In order to use the histogram method efficiently,
we divided the energy range $E<0$ by 10 000 bins. We have verified that
we obtain the same results, with our precision, for 30 000 bins.

The critical slowing down in a first order transition is greater than
in a second order transition because of energy barriers, and thus the 
time of the simulation, to go from one state to another grows 
exponentially with the size of the lattice. For this reason we restricted our
simulations to systems not too large to have good enough statistics.
In each simulation, at least 2 millions (3 millions for the greater sizes)
measurements were made after enough
Metropolis updating (500 000) were carried out to reach equilibration.

For each temperature T we calculate the following quantities
\begin{eqnarray}
\label {titi1}
C&=&\frac {(<E^{2}>-<E>^{2})}{Nk_{B}T^{2}} \ ,\\
\chi &=&\frac {N(<M^{2}>-<M>^{2})}{k_{B}T} \ ,\\
\chi_{\kappa} &=&\frac {N(<\kappa^{2}>-<\kappa>^{2})}{k_{B}T} \ ,\\
\label {titi3}
V&=&1-\frac {<E^{4}>}{3<E^{2}>^{2}} \ , 
\end{eqnarray}
where $M$ is the order parameter,
$C$ the specific heat per site,
$\chi $ the magnetic susceptibility per site,
$V$ the fourth order energy cumulant,
$<...>$ means the thermal average.

The finite size scaling (FSS) for a first order 
transition has been  extensively studied 
\cite{Privman,Binder2,Billoire2}.
A first order transition should be identified by the following properties:
\begin{itemize}
\item[a)] $P_T(E)$ has a double peak.
\item[b)] The maximum of the specific heat $C$ and the susceptibilities $\chi$
and $\chi_{\kappa}$ are proportional to the volume $L^d$. 
\item[c)] The minimum of the fourth order energy cumulant $V$ varies as:
\begin{eqnarray}
\label{tyty3}
V&=& V^* + b\, L^{-d} \ , 
\end{eqnarray}
where $V^*$ is different from 2/3.
\item[d)] The temperatures $T(L)$ at which the quantities 
$C$, $\chi$ or $\chi_{\kappa}$ have a maximum
should vary as:
\begin{eqnarray}
\label{tyty4}
T(L) = T_c + a\, L^{-d}. 
\end{eqnarray}
\end{itemize}
All this conditions will be verified for our systems.
 
\section{Results}     
In Fig. 4 the specific heat $C$ of the STAR model is plotted
  as function of the temperature 
for various sizes (we note that the maximum
is 30 times as large as the usual value of STA which is a sign
of a first order transition). 
The value of the maximum as function of the volume is shown 
in Fig. 5 for $C$, $\chi$ and $\chi_{\kappa}$. 
We note that the maxima vary like the volume 
except for the smaller sizes 
where further corrections are important. 

In Fig. 6 the same quantities are shown for the Stiefel model.
In all cases the maxima vary for the greater sizes proportional to the volume
as they should for a first order transition.

In Fig. 7 and 8 we have plotted $V$ as function of $T$ for different sizes
$L$ for the STAR and the Stiefel model respectively. We can see that 
$V$ does not tend to 2/3 (for a second order transition) but to 
a value $V^*<2/3$.
This value is calculated by fiting the minimum with (\ref{tyty3}). As result
we yield for the STAR model
\begin{eqnarray}
V^* = 0.652(2)
\end{eqnarray}
and for the Stiefel model
\begin{eqnarray}
V^* = 0.625(3) \ .
\end{eqnarray}

The Fig. 9 and 10 show the energy distribution for different sizes at 
different temperatures for the STAR and Stiefel model.
The double peaks observed, even for a very small $L$, indicate a strong
first order transition. With increasing sizes, these two peaks are 
separated by a region of zero probability, indicating a discontinuity of the
energy at the transition. We estimate the correlation length $\xi_0$
by $1/3$ of the first size where the two peaks are well separated 
by a region of zero probability. This is our 
estimate of the distance needed for two 
phases to coexist.
This method yields the correct 
answer in the case of Potts models.
We obtain $\xi_0\sim12$ for the STAR model and
$\xi_0\sim9$ for the Stiefel model (see Fig. 9 and 10).

To obtain the critical temperature we can use (\ref{tyty4}). 
The results are 
\begin{eqnarray}
T_c = 2.2990(5)
\end{eqnarray}
for the STAR model
and
\begin{eqnarray}
T_c = 2.4428(4)
\end{eqnarray}
and for the Stiefel model.
We have sizable   corrections for the small systems.
Comparing the last results with those of Kunz and Zumbach \cite{Kunz}
for the case $V_{2,2}$ similar to ours we agree with their result  
$T_c$=2.445.
 
Our results show clearly the first order transition for 
the STAR and Stiefel model. 
So we confirm the indication given in \cite{Kunz} for the Stiefel model where
in high temperature region $\nu$ was determined not too 
far away from 1/3 which is the value for a first order transition.

\section {Discussion}

We have shown that the STAR and the Stiefel model have first order transitions.
These models are equivalent to the STA in the RG theory,
because the constraint of local rigidity is not relevant
in the transition region. Consequently the phase transition for the
original triangular antiferromagnet STA must also be of first order,
and this result holds generally for all systems with a breakdown of
symmetry of type $O(2)\otimes Z_2$. How can we reconcile our results
with those of the MC of the STA model and the experimental results?

For the MC simulation of the original frustrated spin system
the second order transition is an effect of the finite system size
according to Zumbach's analysis of "almost second order phase transitions"
\cite{Zumbach93}.
The main point of this analysis is that the stable fixed point $F_c$,
known to exist only for the number of components $N>N_c$,
moves into the unphysical complex plane when $N < N_c$. In our case 
$N = 2$, the estimation for $N_c$ is $N_c = 3.91$ \cite{Antonenko 94} and  
second order could never occur.
Nevertheless this complex fixed point has a large basin
of attraction and mimics a behavior of a real fixed point. Only if the system
is very large, i.e. if $L \ge \xi_{0}$, where $\xi_{0}$ 
is the largest correlation length, the transition will appear of first order.
The phenomenon of a crossover between a second order to a first order
transition is not so uncommon. An extreme case is the  Potts model
in two dimensions with $q=5$ components, where the transition is known
to be of first order \cite{Baxter}. The MC gives always a second order
transition with critical exponents of an instable fixed point \cite{Landau}
due to the enormous correlation length $\xi_0$.

If the Hamiltonian is a sum of two terms, one interaction is of the
Heisenberg symmetry and the other favors Ising symmetry:
$H =H_{Heisenberg} + H_{Ising}$ and if $H_{Ising} \ll H_{Heisenberg}$,
one has a crossover between a region of Heisenberg type to Ising behavior
close to $T_c$. If the system size is too small we will only see the region
controlled by the Heisenberg fixed point and therefore obtain the exponents
of the Heisenberg magnet.
In a sense we have the same situation if we replace the Heisenberg
fixed point by the Zumbach fixed point. However, there is 
an essential difference:
for the fixed point in the complex plane, one has to modify the scaling relation
\begin{equation}
\label{tutu1}
\gamma/\nu = 2 - \eta + c
\end{equation}
by a constant $c$ different from zero \cite{Zumbach93}. If the fixed point
is real $c$ must be zero.

We can use this relation as a criterion for real or complex fixed points.
In three dimensions $\eta$ is usually small, that is $\sim 0.03$ (see table
\ref{table6}), but it must be positive \cite{Patashinskii}
and therefore $\gamma/\nu\le2$ for 
a genuine second order phase transition.
If the ratio $\gamma/\nu > 2$ the fixed point must
be complex.  The correction $c$ in the scaling law (\ref{tutu1})
will depend on the distance of the fixed point from real space.
Therefore one expects that $c$ will be greater for the $XY \ (N=2)$ case
than for the Heisenberg $(N=3)$ case. For frustrated Heisenberg systems
it will be difficult to find out whether $\eta_{eff}=\eta-c$ is negative.

With the relation $\gamma/\nu = 2 - \eta $ we obtain $\eta=-0.06$
using the results of MC simulation \cite{PlumerXY} and $\eta=-0.16$
from experimental values of Ho \cite{Thurston1}.
Thus the fixed point is in the complex plan and the second order transition
observed in the XY systems has only an ``almost second order'' character.

The effect of imposing local rigidity obviously forces the system to stay
away from the region of influence of the complex fixed point $F_c$ and
thus permits to "see" the true first order behavior.
Introducing larger coupling constants for inter--plane interactions
than for intra--plane interactions \cite{Plumer97}
seems to have a similar effect as the local rigidity constraint.

A further remark concerning MC simulations and the fixed point $F_c$
in the complex plane;
Zumbach \cite{Zumbach93} has shown that the finite size scaling in this
case does not hold.
The FSS ansatz for the free energy should be replaced by
\begin{equation}
f = L^{-d} g[L/\xi,c_2\mbox{ln}(L)] \ ,
\end{equation}
where $g$ is a function of $L/\xi$, but also of ln$L$.
The constant $c_2$ is proportional
to the constant $c$ of (\ref{tutu1}).
This constant is small and therefore the correction to FSS.
Indeed if we take for the true value of $\eta$ the value of the ferromagnetic
case $\eta\sim 0.03$
(see table \ref{table6}) the value of $c$ will be at most
equal to $c\sim0.03+0.06\sim0.1$ (see discussion above)
and if we compare with $2-\eta$ in (\ref{tutu1}) this gives an error of
5\thinspace \%.
However, small but not negligible corrections to the standard FSS
could explain the differences in MC simulations obtained with different methods
(see table \ref{table5}) and also some of the differences in the experimental
values (tables \ref{table1}-\ref{table4}).

We will discuss now the experiments in the light of the concepts used.
In order to see the first order region the temperature resolution
is the limiting factor not the finite size. However, they are linked through
$t_0 \propto \xi_{0}^{-1/\nu}$ with $\nu \sim 0.5$ found by MC and
$\xi_{0}$ depending on the materials studied.
The temperature distance 
$t_0 = (T - T_c)/T_c$ could be too small to be measurable.

The experiments on CsMnBr$_3$
\cite{Mason2,Mason3,Ajiro,Gaulin,Mason1,Kadowaki1,Wang1,Weber,Goto,Deutschmann,Collins1},
RbMnBr$_3$ \cite{Kato1,Kato2} and CsVBr$_3$ \cite{Tanaka} (\ref{table1})
give exponents compatible with those of MC on STA and a second order
transition (see table \ref{table1} and \ref{table5}).
We can interpret this result by the fact that the
systems are under the influence of a complex fixed point and
$t_{0}$ is too small to observe a first order transition.\\
The case CsCuCl$_3$ \cite{Weber2} (table \ref{table1})
is different since the authors observe a crossover
from a second order region with exponents compatible with MC on STA for
$10^{-3} < t < 5.10^{-2}$ to a region of first order transition for
$5.10^{-5} < t < 5.10^{-3}$.
For $t < t_0 \approx 10^{-3}$ one seems to observe the true first order
region.

Helimagnetic rare earth metals are more complicated as already discussed
in the introduction (see also tables \ref{table2}-\ref{table4}).
The results compatible with those of the MC on STA for Ho
\cite{Jayasuriya1,Thurston1,Gaulin1} (table \ref{table2}),
Dy \cite{Gaulin1,Lederman,Jayasuriya} (table \ref{table3}) and
Tb \cite{Jayasuriya3,Dietrich,Tang1,Hirota,Tang2} (table \ref{table4})
can be interpreted as before: the systems
are under the influence of $F_c$. The first order transition
for Ho \cite{Tindall1} and Dy \cite{Zochowski,Astrom} is due
to the fact that the measurements were done in the first order region
near the
critical temperature. The values of the exponent $\beta\sim0.39$
in the case of Ho and Dy (see table \ref{table2}-\ref{table3})
are not compatible with those found by MC
($\beta\sim0.25$).
This fact can be explained by the
presence of a second length scale in the critical fluctuations near $T_c$
related to random strain fields which are localized at or near the sample
surface \cite{Thurston1}. Thus the critical exponent $\beta$ measured
is of this second length. However the result of $\beta$
for Tb (table \ref{table4}) shows values compatible with MC but it
has been proved that this second length is present also in Tb
\cite{Hirota}.
Further measurements to determine $\beta$ should help in the interpretation.

We have tried to give a general picture of the phase transition
of frustrated XY spins where the breakdown of symmetry is of type
$O(2)\otimes Z_2$. We have shown that this transition is first order but
usually not seen because of the presence of a fixed point
in the complex plane. One way to observe that the behavior is really
driven by such a fixed point is the existence of a negative value for
$\eta$ in the Monte Carlo simulations and experimental systems.
The method used here is to impose local rigidity. This constraint does not
change the behavior of the system but permits the system to stay outside
the region of influence of the complex fixed point.
So we can rely on the
renormalization group study and the true behavior is first order. 
There is no "new chiral
universality class" in strict sense for 
our system ($N=2$). Another possibility discussed in the
literature is that the transition is influenced by the presence of topological
defaults which are not visible in the continuum formulation
of the RG (for the presence of topological defaults in Stiefel
model see \cite{Kunz}).

From our experience with the frustrated $XY$--model we conclude
that the true first order transition for the frustrated Heisenberg model
cannot be reached in MC simulations.
The experimental situation should be similar. However,
due to presence of the anisotropies we never reach the Heisenberg first order
region but will have a crossover to the Ising or $XY$ region \cite{Loison 98}.

\section {Acknowledgments}
This work was supported by the Alexander von Humboldt Foundation.
One of the authors (D.L.) is grateful to Professors B. Delamotte, H.T. Diep 
and A. Dobry for discussions,
and for A.I. Sokolov for the reference to the proof of $\eta \ge 0$.

\onecolumn

\begin{table}[t]
\begin{center}
\begin{tabular}{c|c|c|c|c|c|c}
\hspace{-10pt}
\begin{tabular}{c}Crystal\end{tabular}
\hspace{-10pt}
&
\begin{tabular}{c}method\end{tabular}
\hspace{-10pt}
&
\begin{tabular}{c}ref.\end{tabular}
\hspace{-10pt}
&
\begin{tabular}{c}$\alpha$\end{tabular}
\hspace{-10pt}
&
\begin{tabular}{c}$\beta$\end{tabular}
\hspace{-10pt}
&
\begin{tabular}{c}$\gamma$\end{tabular}
\hspace{-10pt}
&
\begin{tabular}{c}$\nu$\end{tabular}
\hspace{-10pt}
\\
\hline
CsMnBr$_3$&Neutron&\cite{Mason3}&&0.22(2)&&
\hspace{-10pt}
\\
\hline
CsMnBr$_3$&Neutron&\cite{Ajiro}&&0.25(1)&&
\hspace{-10pt}
\\
\hline
CsMnBr$_3$&Neutron&\cite{Gaulin}&&0.24(2)&&
\hspace{-10pt}
\\
\hline
CsMnBr$_3$&Neutron&\cite{Mason1}&&0.21(2)&1.01(8)&0.54(3)
\hspace{-10pt}
\\
\hline
CsMnBr$_3$&Neutron&\cite{Kadowaki1}&&&1.10(5)&0.57(3)
\hspace{-10pt}
\\
\hline
CsMnBr$_3$&Calorimetry&\cite{Wang1}&0.39(9)&&&
\hspace{-10pt}
\\
\hline
CsMnBr$_3$&Calorimetry&\cite{Deutschmann}&0.40(5)&&&
\hspace{-10pt}
\\
\hline
RbMnBr$_3$&Neutron&\cite{Kato1}&&0.28(2)&&
\hspace{-10pt}
\\
\hline
RbMnBr$_3$&Calorimetry&\cite{Kato2}&0.22-0.42&&&
\hspace{-10pt}
\\
\hline
CsCuCl$_3$&Neutron&\cite{Adachi}&&0.25(2)&&
\hspace{-10pt}
\\
\hline
CsCuCl$_3$&Calorimetry&\cite{Weber2}&0.35(5) if $10^{-3} < t < 5.10^{-2}$&&&
\hspace{-10pt}
\\
\hline
CsCuCl$_3$&Calorimetry&\cite{Weber2}&$>0.6$ if $5.10^{-5} < t < 5.10^{-3}$&&&
\hspace{-10pt}
\\
\end{tabular}
\end{center}
\caption{\protect\label{table1}
Experimental values of critical exponents for compound AXB$_3$ 
}
\end{table}

\vspace{2cm}

\begin{table}[t]
\begin{center}
\begin{tabular}{c|c|c|c|c|c}
\hspace{-10pt}
\begin{tabular}{c}Crystal\end{tabular}
\hspace{-10pt}
&
\begin{tabular}{c}ref.\end{tabular}
\hspace{-10pt}
&
\begin{tabular}{c}$\alpha$\end{tabular}
\hspace{-10pt}
&
\begin{tabular}{c}$\beta$\end{tabular}
\hspace{-10pt}
&
\begin{tabular}{c}$\gamma$\end{tabular}
\hspace{-10pt}
&
\begin{tabular}{c}$\nu$\end{tabular}
\hspace{-10pt}
\\
\hline
Thermal expansion&\cite{Tindall1}&1$^{st}$ order $\ $
\hspace{-10pt}
\\
\hline
Calorimetry&\cite{Jayasuriya1}&0.27(2)&&&
\hspace{-10pt}
\\
\hline
Calorimetry&\cite{Wang1}&0.10-0.22&&&
\hspace{-10pt}
\\
\hline
Neutron&\cite{Gaulin1}&&&1.14(10)&0.57(4)
\hspace{-10pt}
\\
\hline
Neutron&\cite{Thurston1}&&0.3(1)&1.24(15)&0.54(4)
\hspace{-10pt}
\\
\hline
Neutron&\cite{DuPlessis1}&&0.39(2)&&
\hspace{-10pt}
\\
\hline
Neutron&\cite{Eckert1}&&0.39(4)&&
\hspace{-10pt}
\\
\hline
X-ray&\cite{Thurston1}&&0.37(1)&&
\hspace{-10pt}
\\
\hline
X-ray&\cite{Helgesen}&&0.39(4)&&
\hspace{-10pt}
\\
\end{tabular}
\end{center}
\caption{\protect\label{table2}
Experimental values of critical exponents for Holmium  
}
\end{table}

\vspace{2cm}

\begin{table}[t]
\begin{center}
\begin{tabular}{c|c|c|c|c|c}
\hspace{-10pt}
\begin{tabular}{c}method\end{tabular}
\hspace{-10pt}
&
\begin{tabular}{c}ref.\end{tabular}
\hspace{-10pt}
&
\begin{tabular}{c}$\alpha$\end{tabular}
\hspace{-10pt}
&
\begin{tabular}{c}$\beta$\end{tabular}
\hspace{-10pt}
&
\begin{tabular}{c}$\gamma$\end{tabular}
\hspace{-10pt}
&
\begin{tabular}{c}$\nu$\end{tabular}
\hspace{-10pt}
\\
\hline
Thermal expansion&\cite{Zochowski}&1$^{st} $ order $\ $
\hspace{-10pt}
\\
\hline
Calorimetry&\cite{Astrom}&1$^{st}$ order $\ $
\hspace{-10pt}
\\
\hline
Calorimetry&\cite{Lederman}&0.18(8)&&&
\hspace{-10pt}
\\
\hline
Calorimetry&\cite{Jayasuriya}&0.24(2)&&&
\hspace{-10pt}
\\
\hline
Neutron&\cite{Gaulin1}&&&1.05(7)&0.57(5)
\hspace{-10pt}
\\
\hline
Neutron&\cite{DuPlessis1}&&0.38(2)&&
\hspace{-10pt}
\\
\hline
Neutron&\cite{DuPlessis2}&&0.38(3)&&
\hspace{-10pt}
\\
\hline
Neutron&\cite{Brits}&&0.39(1)&&
\hspace{-10pt}
\\
\hline
M\"ossbauer&\cite{Loh}&&0.335(10)&&
\hspace{-10pt}
\\
\end{tabular}
\end{center}
\caption{\protect\label{table3}
Experimental values of critical exponents for dysprosium 
}
\end{table}

\vspace{2cm}

\begin{table}[t]
\begin{center}
\begin{tabular}{c|c|c|c|c|c}
\hspace{-10pt}
\begin{tabular}{c}method\end{tabular}
\hspace{-10pt}
&
\begin{tabular}{c}ref.\end{tabular}
\hspace{-10pt}
&
\begin{tabular}{c}$\alpha$\end{tabular}
\hspace{-10pt}
&
\begin{tabular}{c}$\beta$\end{tabular}
\hspace{-10pt}
&
\begin{tabular}{c}$\gamma \ \ \  $\end{tabular}
\hspace{-10pt}
&
\begin{tabular}{c}$\nu$\end{tabular}
\hspace{-10pt}
\\
\hline
Calorimetry&\cite{Jayasuriya3}&0.20(3)&&&
\hspace{-10pt}
\\
\hline
Neutron&\cite{Dietrich}&&0.25(1)&$\ $&
\hspace{-10pt}
\\
\hline
Neutron&\cite{Tang1}&&0.23(4)&&
\hspace{-10pt}
\\
\hline
Neutron&\cite{Hirota}&&&&0.53
\hspace{-10pt}
\\
\hline
X-ray&\cite{Tang2}&&0.21(2)&&
\hspace{-10pt}
\\
\end{tabular}
\end{center}
\caption{\protect\label{table4}
Experimental values of critical exponents for Terbium 
}
\end{table}

\vspace{2cm}

\begin{table}[t]
\begin{center}
\begin{tabular}{c|c|c|c|c|c|c|c|c}
\hspace{-10pt}
\begin{tabular}{c}ref.\end{tabular}
\hspace{-10pt}
&
\begin{tabular}{c}$\alpha$\end{tabular}
\hspace{-10pt}
&
\begin{tabular}{c}$\beta$\end{tabular}
\hspace{-10pt}
&
\begin{tabular}{c}$\gamma$\end{tabular}
\hspace{-10pt}
&
\begin{tabular}{c}$\nu$\end{tabular}
\hspace{-10pt}
&
\begin{tabular}{c}$\eta^1$\end{tabular}
\hspace{-10pt}
&
\begin{tabular}{c}$\beta_{\kappa}$\end{tabular}
\hspace{-10pt}
&
\begin{tabular}{c}$\gamma_{\kappa}$\end{tabular}
\hspace{-10pt}
&
\begin{tabular}{c}$\nu_{\kappa}$\end{tabular}
\hspace{-10pt}
\\
\hline
\cite{Kawa92}&0.34(6)&0.253(10)&1.13(5)&0.54(2)&-0.09(8)&0.45(2)&0.77(5)&0.55(2)
\hspace{-10pt}
\\
\hline
\cite{PlumerXY}&0.46(10)&0.24(2)&1.03(4)&0.50(1)&-0.06(4)&0.38(2)&0.90(9)&0.55(1) 
\hspace{-10pt}
\\
\hline
\cite{Loison 96}&0.43(10)&&&0.48(2)&&&&                       
\hspace{-10pt}
\\
\end{tabular}
\end{center}
\caption{\protect\label{table5}
Critical exponents by Monte Carlo for $O(2) \otimes Z_2$. 
$^1$calculated by $\gamma/\nu=2-\eta$. The first result
{\protect \cite{Kawa92}} comes from a study
at high and low temperature and uses of FSS. The second 
{\protect \cite{PlumerXY}} uses the Binder 
parameter {\protect \cite{Binder}} to find $T_c$ and uses the FSS,  the third
{\protect \cite{Loison 96}} uses the maxima in FSS region. The results
$\beta_{\kappa}$, $\gamma_{\kappa}$, $\nu_{\kappa}$ are the exponents
for the chirality. 
}
\end{table}

\vspace{2cm}

\begin{table}[t]
\begin{center}
\begin{tabular}{c|c|c|c|c|c}
\hspace{-10pt}
\begin{tabular}{c}symmetry $\ $ \end{tabular}
\hspace{-10pt}
&
\begin{tabular}{c}$\alpha$\end{tabular}
\hspace{-10pt}
&
\begin{tabular}{c}$\beta$\end{tabular}
\hspace{-10pt}
&
\begin{tabular}{c}$\gamma$\end{tabular}
\hspace{-10pt}
&
\begin{tabular}{c}$\nu$\end{tabular}
\hspace{-10pt}
&
\begin{tabular}{c}$\eta$\end{tabular}
\hspace{-10pt}
\\
\hline
$Z_2$&0.107&0.327&1.239&0.631&0.038
\hspace{-10pt}
\\
\hline
$O(2)$&-0.010&0.348&1.315&0.670&0.039
\hspace{-10pt}
\\
\hline
$O(3)$&-0.117&0.366&1.386&0.706&0.038
\hspace{-10pt}
\\
\hline
$O(4)$&-0.213&0.382&1.449&0.738&0.036 
\hspace{-10pt}
\\
\hline
$O(5)$&-0.297&0.396&1.506&0.766&0.034 
\hspace{-10pt}
\\
\hline
$O(6)$&-0.370&0.407&1.556&0.790&0.031 
\hspace{-10pt}
\end{tabular}
\end{center}
\caption{\protect\label{table6}
Critical exponents for the ferromagnetic systems calculated by RG
{\protect \cite{Antonenko1}}.
}
\end{table}

\vspace{2cm}

\twocolumn

\begin{figure}[tbp]
\vskip 1cm
\centerline{
\label{titia}
\epsfxsize = \hsize
\epsfbox{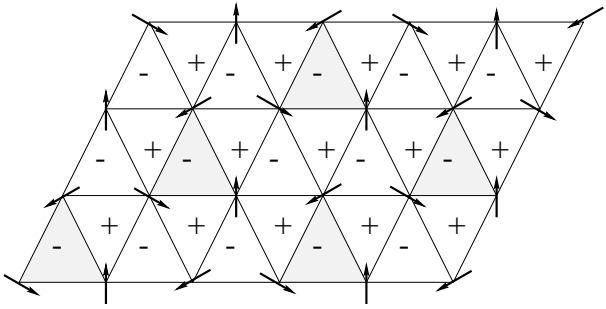} }
\vskip 1cm
\caption{
Ground state configuration for the STA and the STAR model.
The chirality of each triangle is indicated by $+$ or $-$. The other ground
state configuration with opposite chirality is obtained by reversing all spins.
The supertriangles are dark.
}
\end{figure}

\vspace{2cm}

\begin{figure}
\vskip 1cm
\centerline{
\label{titib}
\epsfxsize = \hsize
\epsfbox{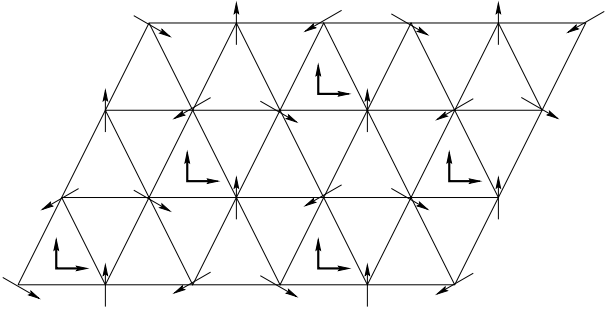} }
\vskip 1cm
\caption{
The dihedral are drawn at the center of each elementary supertriangle.
}
\end{figure}

\vspace{2cm}

\begin{figure}
\vskip 1cm
\centerline{
\label{titic}
\epsfxsize = \hsize
\epsfbox{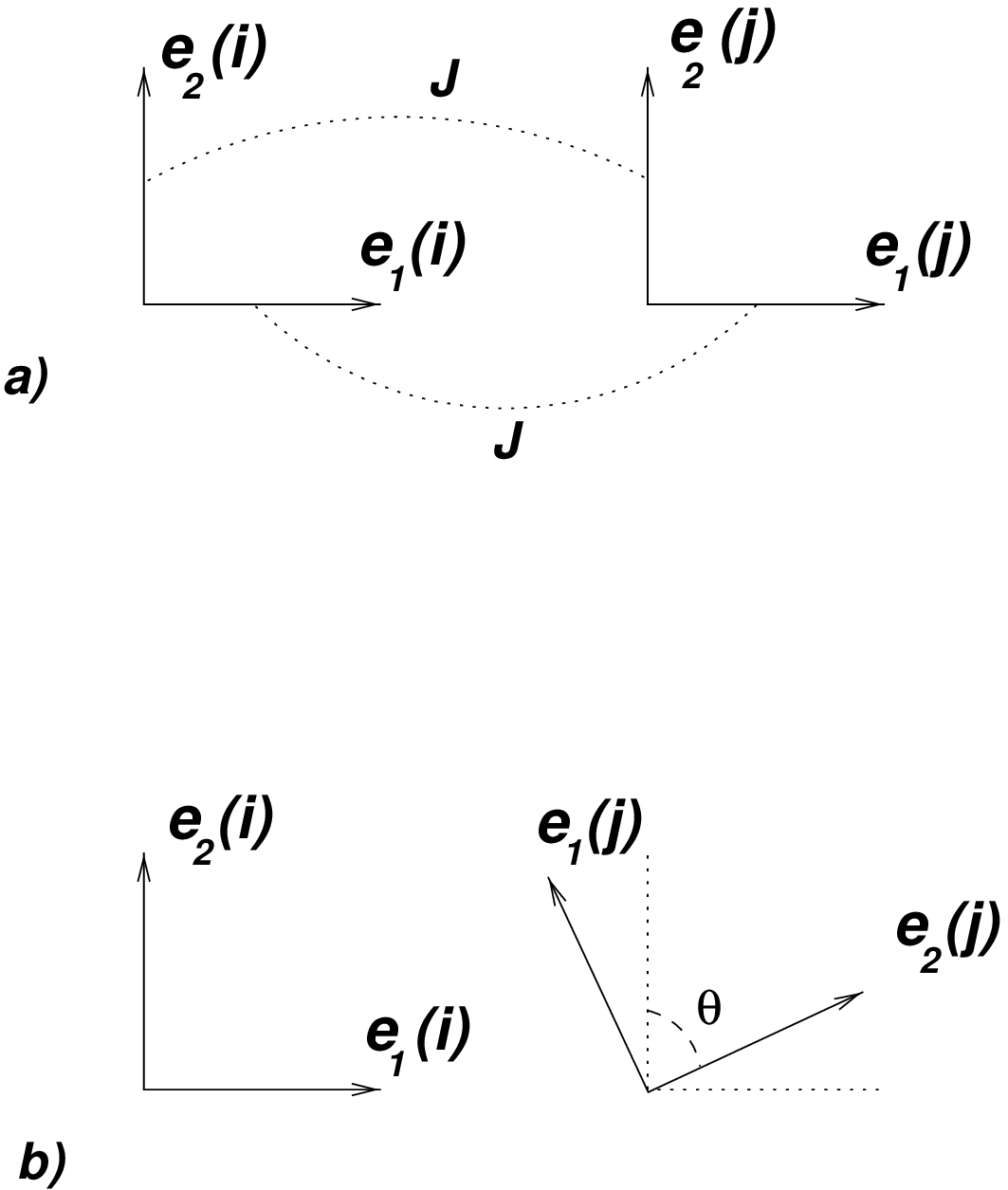} }
\vskip 1cm
\caption{
a. Dihedral and their interaction. ${\bf e}_1(i)$ interacts only with 
${\bf e}_1(j)$ not with ${\bf e}_2(j)$ which interacts with ${\bf e}_2(i)$.
b. Two dihedral with opposite chirality. The energy of the interaction
is equal to $\cos(\pi-\theta)+\cos(\theta)=0$.
}
\end{figure}

\vspace{2cm}

\begin{figure}
\vskip 1cm
\centerline{
\label{titid}
\epsfxsize = \hsize
\epsfbox{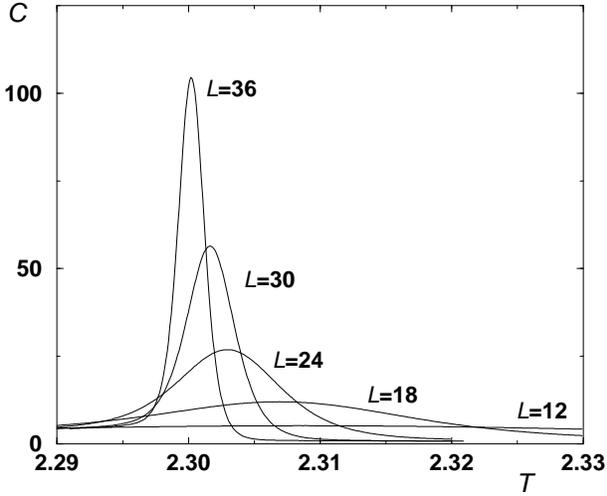} }
\vskip 1cm
\caption{
Specific heat for the STAR model for various sizes as function of
temperature. 
}
\end{figure}

\vspace{2cm}

\begin{figure}
\vskip 1cm
\centerline{
\label{titie}
\epsfxsize = \hsize
\epsfbox{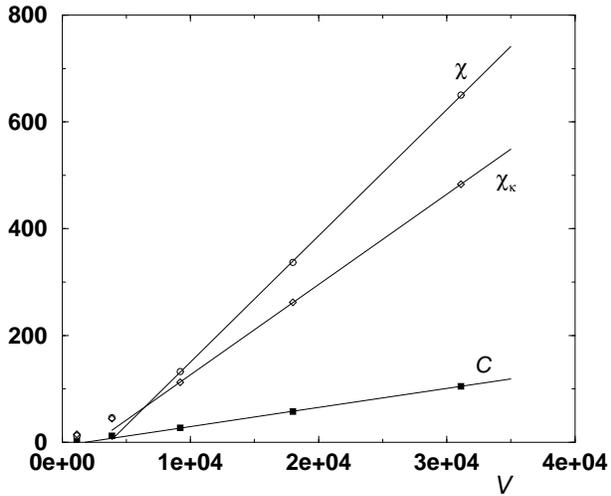} }
\vskip 1cm
\caption{
Maxima of the specific heat C, the susceptibility 
$\chi$ (magnetization) and $\chi_{\kappa}$ (chirality) in function 
of the volume $V=L^2\cdot L_z$ for the STAR model.
}
\end{figure}

\vspace{2cm}

\begin{figure}
\vskip 1cm
\centerline{
\label{titif}
\epsfxsize = \hsize
\epsfbox{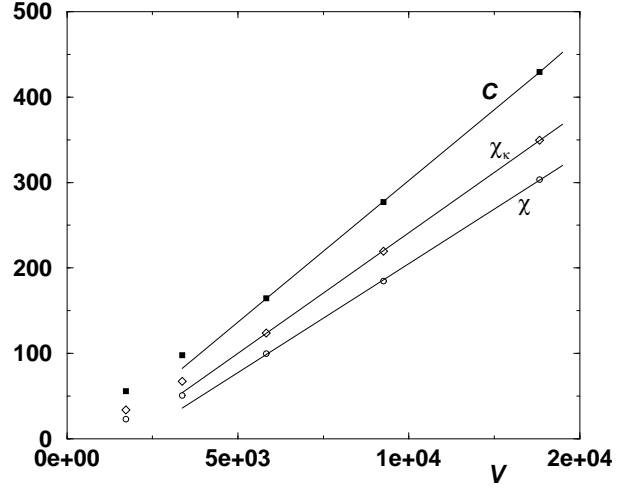} }
\vskip 1cm
\caption{
Maxima of the specific heat C, the susceptibility 
$\chi$ (magnetization) and $\chi_{\kappa}$ (chirality) in function 
of the volume $V=L^3$ for the Stiefel model.
}
\end{figure}

\vspace{2cm}

\begin{figure}
\vskip 1cm
\centerline{
\label{titig}
\epsfxsize = \hsize
\epsfbox{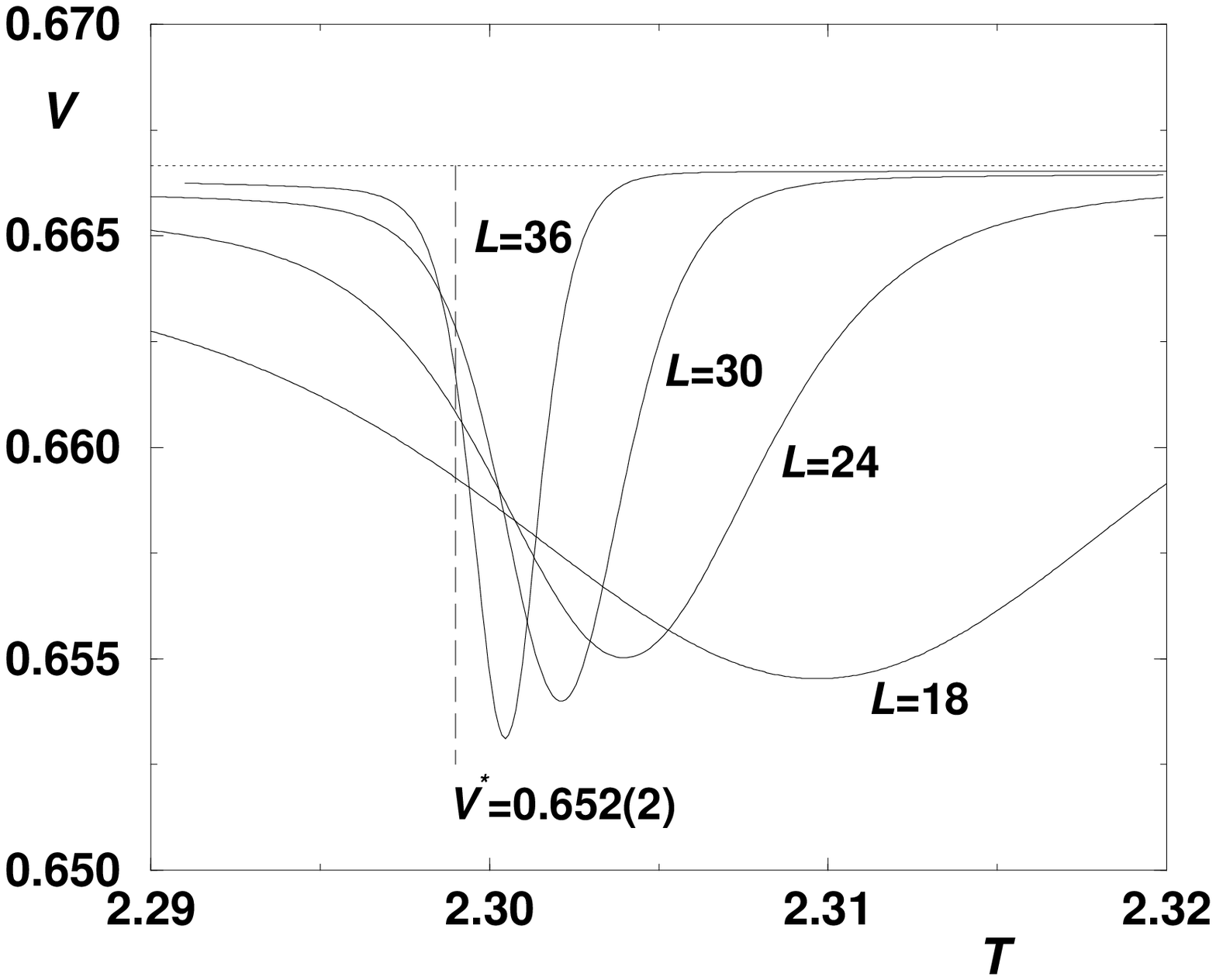} }
\vskip 1cm
\caption{
$V$ as function of temperature for different sizes for the STAR model. 
The \dots line indicate the value of $V^*=2/3$ 
for the second order transition. 
The broken line is our estimate of $V^*$ at the critical 
temperature $T_c=2.2990$. 
}
\end{figure}

\vspace{2cm}

\begin{figure}
\vskip 1cm
\centerline{
\label{titih}
\epsfxsize = \hsize
\epsfbox{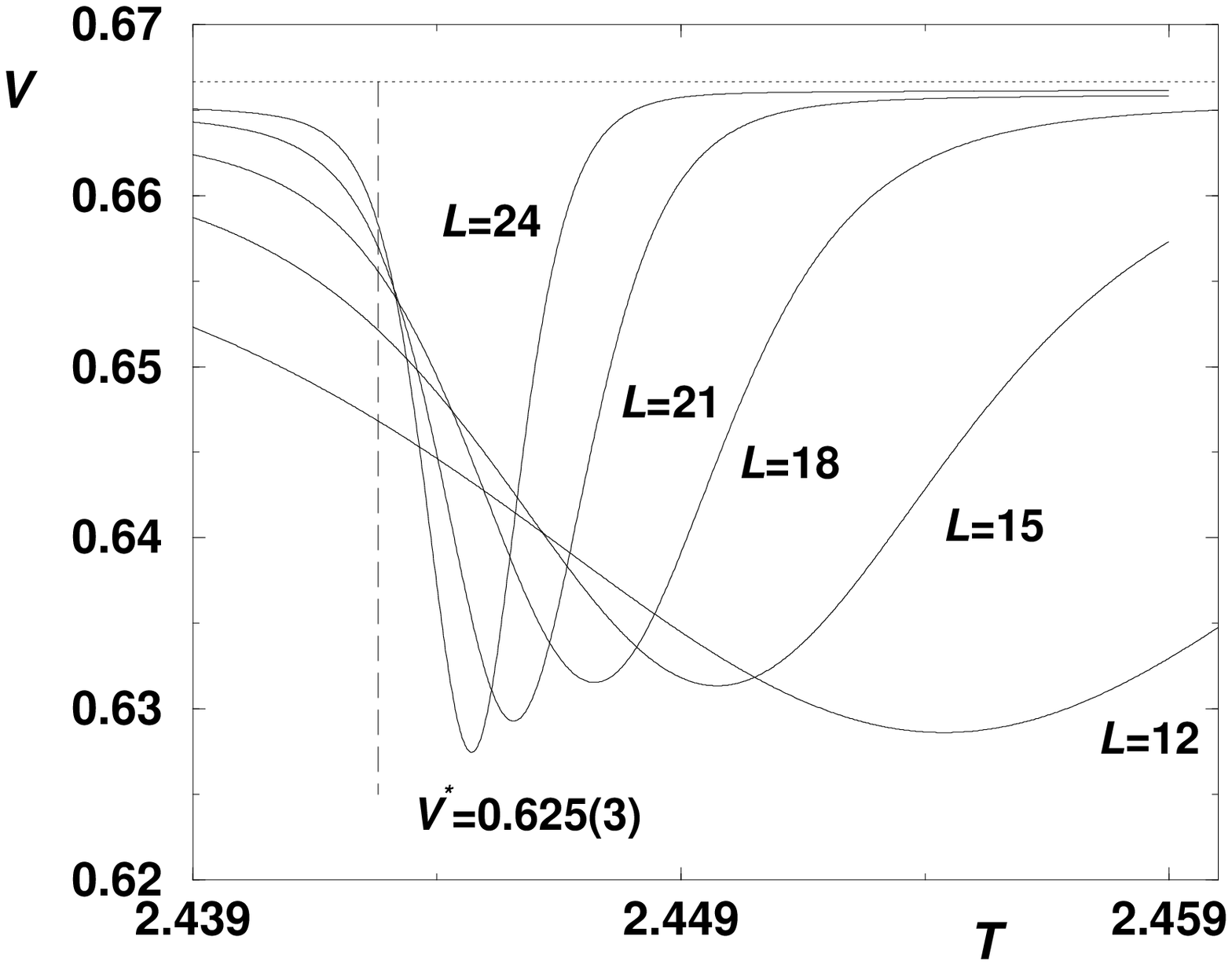} }
\vskip 1cm
\caption{
$V$ as function of the temperature for different sizes for the Stiefel model. 
The \dots line indicate the value of $V^*=2/3$ for the second order transition. 
The broken line is our estimate of $V^*$ at the critical 
temperature $T_c=2.4428$.
}
\end{figure}

\vspace{2cm}

\begin{figure}
\vskip 1cm
\centerline{
\label{titii}
\epsfxsize = \hsize
\epsfbox{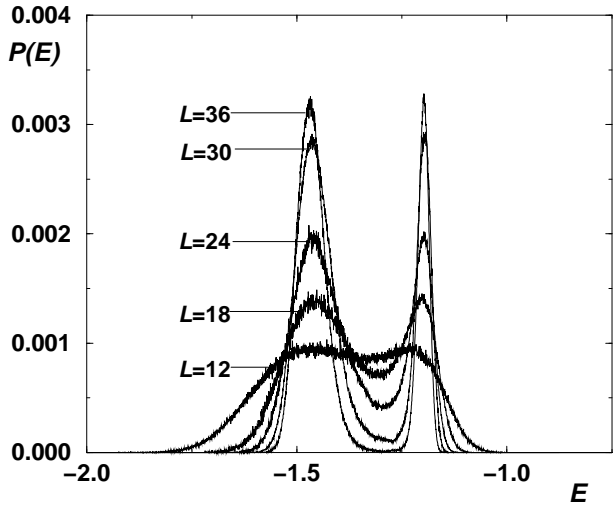} }
\vskip 1cm
\caption{
Energy histogram $P(E)$ as function of energy per site $E$ for the STAR 
model for various sizes $L$ at different temperatures of simulation: 
$T_{12}=2.3136$, $T_{18}=2.3065$, 
$T_{24}=2.3020$, $T_{30}=2.30085$, $T_{36}=2.29968$.
}
\end{figure}

\vspace{2cm}

\begin{figure}
\vskip 1cm
\centerline{
\label{titij}
\epsfxsize = \hsize
\epsfbox{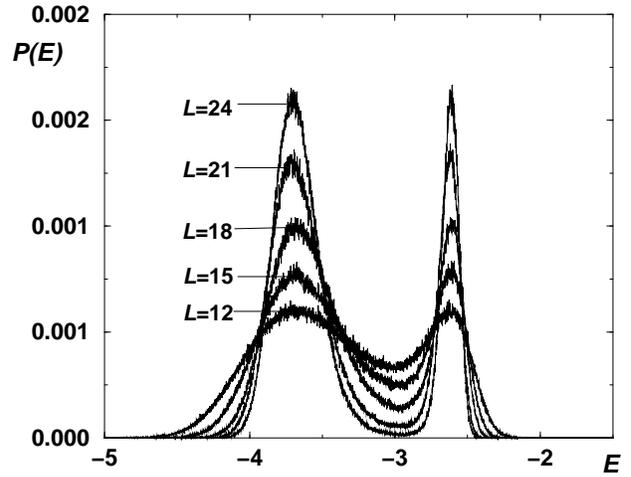} }
\vskip 1cm
\caption{
Energy histogram $P(E)$ as function of energy per site $E$ for the Stiefel
model for various sizes $L$ at different temperatures of simulation:
$T_{12}=2.4495$, $T_{15}=2.4468$,
$T_{18}=2.44525$, $T_{21}=2.44425$, $T_{24}=2.44377$.
}
\end{figure}

\vspace{2cm}

\end{document}